# Disk Galaxies at Z=0 and at High Redshift: An Explanation of the Observed Evolution of Damped Lyα Absorption Systems

Guinevere Kauffmann

*Max-Planck Institut für Astrophysik, D-85740 Garching, Germany*

## Abstract

The analysis of disk formation in this paper is based on the White & Rees (1977) picture, in which disk galaxies form by continuous cooling and accretion of gas within a merging hierarchy of dark matter halos. A simple Kennicutt law of star formation for disks, based on a single-fluid gravitational stability model, is introduced. Since the gas supply in the disk is regulated by infall from the surrounding halo, the gas is always maintained at a critical threshold surface density $\Sigma_c$, where $\Sigma_c \propto V_c/R$. Chemical enrichment of the disks occurs when the surrounding hot halo gas is enriched with heavy elements ejected during supernova explosions. This gas then cools onto the disk producing a new generation of metal-rich stars.

We show that models of this type can reproduce many of the observed properties of a typical spiral galaxy like the Milky Way, including its gas and stellar surface density profiles and the observed relationship between the ages and metallicities of solar neighbourhood stars. In particular, we find that the rapid early enrichment predicted by our model solves the classic G-dwarf problem. In addition, we are able to account for some of the global trends in the properties of disk galaxies, such the observed relation between galaxy luminosity, metallicity and gas content.

We then use our models to make inferences about the properties of disk galaxies at high redshift. Because the overall mass distribution in the universe shifts to smaller halos at higher redshifts, and these smaller halos contain less luminous, more gas-rich galaxies, we find that the total neutral hydrogen density $\Omega(HI)$ increases at higher z. The predicted increase is mild, but is roughly consistent with the latest derivation of $\Omega(HI)$ as a function of $z$ by Storrie-Lombardi & MacMahon (1995). Our models are also able to account for some of the other trends seen in the high-redshift data, including the increase in the number of high column-density systems at high redshift, as well as the metallicity distribution of damped Lyα systems at $z \sim 2 - 3$. Finally, one completely general prediction of these models is that at high redshift, the galaxies that produce the damped Lyα absorption will typically be smaller, more compact and less luminous than disk galaxies at the present epoch.

# 1 Introduction

The damped Ly$\alpha$ systems are the rarest of the different classes of quasar absorption lines, yet they have attracted considerable attention because of indications that they are the high-redshift progenitors of present-day galaxies. The most persuasive evidence comes from the integrated mass of neutral hydrogen contained in these systems at redshifts between 2 and 3, which is comparable to the total mass in stars in galaxies today. In addition, the neutral hydrogen mass density is seen to decrease with redshift, and at the lowest redshifts surveyed so far, is in good agreement with the global density of neutral gas determined from 21 cm studies of nearby galaxies (Rao & Briggs 1993, Fall & Pei 1993). The conventional theoretical picture is one in which pure HI disks are assembled at some high redshift. The gas is then transformed into stars until roughly 90 percent has been "used up" by z=0. Models of this type have been explored in detail by Lanzetta and coworkers (Lanzetta, Wolfe & Turnshek 1995; Wolfe et al 1995). Models taking into account the effects of dust on the inferred column density distribution of the damped Ly$\alpha$ systems have been explored by Fall & Pei (1993) and Pei & Fall (1995).

Although models of this type are certainly illustrative, they are not realistic representations of the formation histories of real disk galaxies. According hierarchical clustering theories, which currently constitute the standard paradigm of structure formation in the universe, galaxies, groups and clusters form continuously through the merging of small subunits to form larger and larger systems. In this scenario, the evolution of the *dark matter component* of the universe is roughly self-similar; only the scale of the collapsed structures changes with time. If it is indeed the evolution in the clustering of the dark matter component that regulates the formation and evolution of galaxies, one might expect that to first order, galaxies at high redshift be rather similar to galaxies today. The main difference would be that the "typical" galaxy be of lower mass and luminosity. Thus if one takes hierarchical clustering theories seriously, the suggestions of Pettini,Boksenberg & Hunstead (1990),York (1988) and Tyson(1988) that the damped systems may be dominated by a population of dwarf galaxies with properties similar to dwarf galaxies today may warrant re-examination.

In this paper, we explore predictions for the gas properties of high redshift galaxies in cold dark matter (CDM) models, in which disk galaxies form as gas cools and forms stars at the centres of dark matter halos. In the model, disk gas is continuously replenished as a result of infall from the surrounding hot halo. This type of model requires a substantial source of energy input to keep *all the gas* from cooling off and forming dense lumps in small halos at high redshift, where the cooling times are short. Possibilities include energy injection by supernova explosions that occur soon after the first generation of stars begin to form in a galaxy. Alternatively, gas may be prevented from cooling off at all in small systems at high redshift by the presence of a photo-ionizing background of UV-radiation produced by quasars (Efstathiou 1992). Continuous infall models appear to be necessary in order to explain the observed sizes of disk galaxies today. Simulations that do not incorporate any heating processes produce cold gas disks with scale radii much too small to be compatible with observations (Navarro & White 1994; Steinmetz 1995). These authors find that a substantial amount of angular momentum is lost to the dark matter during merging of the small dense lumps of gas that are able to cool off at high redshift. On the other hand, hot gas halos around ordinary spiral galaxies have yet to be observed in the X-ray, where they should be most easily visible.

## 2 Description of the Model

### 2.1 Formation and Merging of Dark Matter Halos

The formation, evolution and merging histories of the dark matter halos in which gas will cool and condense, is specified using a semi-analytic technique developed by Kauffmann & White (1993), based on an extension of the Press-Schechter theory due to Bower (1993) and Bond et al (1993). The original Press-Schechter theory gave the mass distribution of collapsed, virialized objects in the universe as a function of redshift. The formalism was applicable to any set of cosmological initial conditions resulting in the *hierarchical* buildup of structure. With the extended theory, one is able to specify the probability that a halo of mass $M_1$ at redshift $z_1$ will later be incorporated into a halo of mass $M_0$ at $z_0$. Extensive tests of the theory using numerical simulations of gravitational clustering have been carried out and remarkably good agreement has been found (Kauffmann & White 1993, Lacey & Cole 1994).

In our models, we construct Monte Carlo realizations of the merging history of present-day halos of given mass. In this way, we follow not only the evolution in the global mass distribution of halos as a function of redshift, but also the formation and growth of individual halos with time. The interested reader is referred to Kauffmann & White (1993) for further details.

### 2.2 Cooling of Gas

The treatment of gas cooling is based on a model by White & Frenk (1991). A dark matter halo is modelled as an isothermal sphere that is truncated at its virial radius, defined as the radius within which the mean overdensity is 200. In the standard spherical collapse model, each mass shell is assumed to virialize at one half its maximum expansion radius. The virial radius is then related to the circular velocity of the clump and to redshift by

$$r_{vir} = 0.1 H_0^{-1}(1+z)^{-3/2} V_c. \qquad (1)$$

The virial mass may then be written

$$M_{vir} = 0.1 G^{-1} H_0^{-1}(1+z)^{-3/2} V_c^3. \qquad (2)$$

We assume that when a halo forms, the gas initially relaxes to a distribution which exactly parallels that of the dark matter. The gas temperature is then related to the circular velocity of the halo through the equation of hydrostatic equilibrium. The *cooling radius* is defined as the radius within the halo at which the cooling time is equal to the Hubble time. If the cooling radius lies outside the virial radius, we are in the accretion-limited case where all infalling gas cools immediately. If the cooling radius lies inside the virial radius, we model the cooling rate by a simple inflow equation:

$$\dot{M}_{cool} = 4\pi \rho_g(r_{cool}) r_{cool}^2 \frac{dr_{cool}}{dt} \tag{3}$$

The gas that is able to cool will collapse to form a rotationally-supported disk at the centre of the halo. For gas that collapses within the potential of a massive halo while conserving angular momentum, the collapse factor $f_{diss}$ may be written

$$f_{diss} = \left(\frac{R_D}{R_H}\right) \sim \left(\frac{M_H}{M_D}\right)\left(\frac{\lambda_D}{\lambda_H}\right)^2, \tag{4}$$

where $\lambda_D$ is the spin parameter of the disk, observed to lie in the range 0.4-0.5 for real spiral galaxies, and $\lambda_H$ is the spin parameter of the dark halo. N-body simulations show that halos typically have $\lambda_H \sim 0.05 \pm 0.03$ as a result of tidal torquing (Barnes & Efstathiou 1987), so for a disk/halo mass ratio of 0.1, one obtains a collapse factor of 0.1. In practice, because $\lambda_H$ has substantial scatter, one expects a spread in collapse factors. For simplicity, we will assume that gas will collapse to a constant fraction $f_{diss} = 0.1$ of its radius within its parent halo. Gas distributed isothermally within the halo thus forms a disk with surface mass density $\Sigma \propto r^{-1}$ and outer radius $R_D = 0.1 R_H$.

As seen from equation 1, the size of a dark matter halo is proportional to its circular velocity and scales with redshift as $(1+z)^{-3/2}$. Disks are thus built up from the inside out as the mass of their surrounding dark halos grows with time and gas falls in from larger and larger radii.

## 2.3 Star Formation and Feedback in Galactic Disks

In a seminal paper in 1989, Kennicutt showed that the star formation rates and radial profiles of a sample of nearby spiral galaxies could be explained very simply by a combination of a Schmidt power-law rate of star formation at high gas densities and a cut-off in star formation below a certain critical threshold surface density. In this paper, Kennicutt noted that an abrupt decrease in star formation at low gas densities is expected from simple gravitational stability considerations, as first discussed by Toomre (1964). In the case of a thin isothermal gas disk, instability is only expected if the surface density exceeds the critical value

$$\Sigma_c = \frac{\alpha \kappa c}{3.36 G}, \tag{5}$$

where $\alpha$ is a dimensionless constant near unity, $c$ is the velocity dispersion of the gas, and $\kappa$ is the epicyclic frequency given by,

$$\kappa = 1.41 \frac{V}{R}\left(1 + \frac{R}{V}\frac{dV}{dR}\right)^{1/2}. \tag{6}$$

For his sample of 15 galaxies, Kennicutt showed that the outer radii of the HII regions in these galaxies corresponded extremely well to the radii at which their gas surface densities dropped below the critical density. In addition, within the star-forming disks of the galaxies, the ratio of the gas surface density to the critical density was always close to unity, indicating that the gas disks tended to lie near their gravitational stability limit.

In our models, we adopt the simple form of a Kennicutt star formation law. We adopt a constant gas velocity dispersion $c = 6$ km s$^{-1}$ and assume that all disks have flat rotation curves and a rotational velocity equal to the circular velocity of their surrounding dark matter halos. The epicyclic frequency $\kappa$ is then simply given by $V/R$ and the stability condition takes the form

$$\Sigma_{crit}(M_\odot pc^{-2}) = 0.59 V(kms^{-1})/R(kpc). \tag{7}$$

At densities greater than $\Sigma_{crit}$, we adopt a star formation law of the form

$$\dot{M}_*(M_\odot yr^{-1} pc^{-2}) = \beta \Sigma_{gas}/t_{dyn}, \tag{8}$$

where $\beta$ is a free parameter controlling the efficiency of star formation and $t_{dyn}$ is the dynamical time of the disk $t_{dyn} = (V/R_D)^{-1}$. In practice, the resulting gas profiles of the disks in our model depend rather little on $\beta$ because the supply of gas in the disk is regulated on a short timescale by infall of gas from the surrounding halo. The surface density of gas in the disk always tracks the critical density, except at the outer edges where the disk is only just beginning to form (see section 3.1).

To maintain a supply of hot gas in the halo so that continuous infall is able to take place, we postulate that supernova explosions can release enough energy to drive cold gas back into the hot intergalactic medium. For a standard Scalo IMF, the number of supernovae expected per solar mass of stars formed is $\eta_{SN} = 4 \times 10^{-3} M_\odot^{-1}$. The kinetic energy of the ejecta from each supernova is $10^{51}$ ergs. If a fraction $\epsilon$ of this energy is used to reheat cold gas to the virial temperature of the halo, the amount of cold gas lost to the intergalactic medium in time $\Delta t$ may be estimated from simple energy balance arguments as

$$\Delta M_{reheat} = \epsilon \frac{4}{5} \frac{\dot{M}_* \eta_{SN} E_{SN}}{V_c^2} \Delta t \tag{9}$$

Here $\epsilon$ is a free parameter controlling the efficiency of the feedback process. In practice, the value of $\epsilon$ will determine the total amount of gas that is transformed into stars and hence the luminosity of the galaxy, but the gas profiles in the disks are insensitive to changes in this parameter.

## 2.4 A Simple Model for Chemical Enrichment

The continuous infall models described above require substantial heating of halo gas by star formation activity, so it seems plausible that a large amount of processed material could be mixed into the gas out to large radii in the halo. There are at least two observational indications that this kind of process does indeed occur. The [Mg II] absorption-line systems observed at low redshift in quasar spectra are almost always associated with star-forming galaxies (Bergeron 1988; Steidel, Dickinson & Persson 1994). These systems are often seen at several optical diameters away from the galaxy centre, suggesting that halos are enriched to large radius as a consequence of star formation. The second indication is the mean metallicity of the X-ray emitting gas in rich clusters. The total metal contents of the cluster gas and the stars of the cluster galaxies are comparable. This would indicate that a substantial fraction of heavy elements produced by the stars was not retained by the galaxies.

White & Frenk (1991) have explored chemical evolution models of this type. For every solar mass of stars formed in a galaxy, an *effective* yield $y$ of heavy elements is assumed to be ejected and uniformly mixed with the hot halo gas. The mass of metals in stars thus increases as metals are incorporated from the surrounding hot gas as a result of cooling and infall. The metallicity of the gas is increased by stellar ejecta, but decreased by metals lost to stars and accretion of primordial material as the halos grow in mass. Following the techniques outlined in White & Frenk (1991), the effective yield $y$ is taken as a free parameter and its value is constrained by requiring that $L_*$ galaxies be enriched to roughly solar metallicity by the present day (see section 2.6 for more details). This approach is adopted because of the large uncertainties associated with modelling the ejection and mixing processes. The main purpose of the model is to explore *relative changes* in the metallicity of galaxies as a result of differing accretion or star formation histories. We do not claim that it in any way constitutes a detailed or accurate description of the physical processes affecting the intergalactic medium.

## 2.5 Fixing the Free Parameters in the Model

The parameters $\alpha$, $\epsilon$ and $y$, which respectively control the star formation efficiency, the feedback efficiency and the heavy element yield, are constrained by requiring that *on average*, disk galaxies that form within halos of circular velocity 220 km s$^{-1}$ have properties that match those of our own Milky Way galaxy. As discussed previously, the gas surface density profiles of the disks in our model track the critical density. The total gas mass thus does not depend very much on our choice of parameters. For a Milky Way-type galaxy, $M_{tot}(gas) \sim 10^{10} M_\odot$, which is roughly comparable to our Galaxy's measured HI mass of $8 \times 10^9 M_\odot$. The parameter $\epsilon$ fixes the luminosity of the galaxy and we set its value by requiring that the B-band luminosity $L_B$ have a value of $\sim 2 \times 10^{10} L_\odot$ for a Milky Way-type galaxy. Finally, the yield is fixed by requiring that the mean metallicity of the stars in the Milky Way be close to solar.

We restrict ourselves to cold dark matter (CDM) initial conditions with $\Omega = 1$, $H_0 = 50$ km s$^{-1}$ Mpc$^{-1}$ and $\Omega_{baryon} = 0.1$. Whenever not specified, we adopt a normalization with $\sigma_8 = 0.67$ ($b = 1.5$).

# 3 Results of the Model

## 3.1 The Properties of Disk Galaxies at z=0

We will first explore to what extent our models can reproduce the properties of disk galaxies at the present day. The left-hand panel of figure 1 shows the HI and stellar surface density profiles of a disk galaxy residing in a halo with circular velocity $V_c = 220$ km s$^{-1}$. The dotted line is the critical surface density defined in equation 7. As can be seen, the HI surface density tracks the critical density out to a radius of 20 kpc. Beyond this radius, gas is falling in for the first time and there has not yet been time for the surface density to reach the critical density and for star formation to "switch on". This is reflected in the stellar mass density profile, which is truncated abruptly at a radius of 20 kpc. The HI column densities measured through a face-on disk range from a few $\times 10^{21}$ cm$^{-2}$ in the central few kiloparsecs, to $10^{20}$ cm$^{-2}$ at a distance of $\sim$ 30-40 kpc. This agrees rather well with what is observed for real spiral galaxies (Bosma 1981), where N(HI) typically falls below $10^{20}$ cm$^{-2}$ at about 1.5 Holmberg radii. The stellar mass falls off much more steeply than the gas. It is roughly exponential over much of the disk, with a scale length $\sim$ 4 kpc. We thus conclude that our model galaxy is in reasonable agreement with the properties of the Milky Way disk component. In the right-hand panel of figure 1, we show the HI and stellar profiles of a "dwarf" disk galaxy contained in a halo of circular velocity 75 km s$^{-1}$. This galaxy is simply a scaled-down version of its brighter counterpart. The stellar scale length is about 1 kpc, typical of fainter disk systems such as the Magellanic Clouds.

The total HI mass of a galaxy is obtained by integrating the gas surface density over the area of the disk,

$$M_{gas} = 2\pi \int_0^{R_{lim}} \Sigma_{gas}(r) r dr, \qquad (10)$$

where $R_{lim}$ is the outer limit of the disk. Since $\Sigma_{gas}(r) \sim \Sigma_{crit}(r) \propto V_c/r$ and $R_{lim} \propto V_c$, we find that $M_{gas} \propto V_c^2$. Therefore, using equation 2, $M_{gas}/M_{halo} \propto V_c^{-1}$, i.e. smaller halos have higher gas mass fractions. If galaxies obey the B-band Tully-Fisher relation ($L_B \propto V_c^{2.7}$), we obtain a gas-mass/ B-band luminosity relation of the form $M_{gas} \propto L_B^{-0.7}$. As was first realized by Quirk (1972), this offers a natural explanation for why dwarf galaxies appear substantially more gas-rich than bright galaxies.

Star formation histories of Milky Way-type disk galaxies are plotted in figure 2. The star formation rate typically increases from a few tenths of a solar mass per year at redshifts around 10, to values of between 1 and 2 solar masses per year at redshifts around 2-3, by which time a substantial fraction of the mass of the final halo has already been assembled. The star formation rate then remains roughly constant until the present day. It should be noted that at no time does a disk galaxy have very high rates of star formation.

The simple chemical evolution model outlined in section 2.4 enables us to determine the metallicity distribution of stars in a Milky Way-type disk at the present day. Recall that in our model, metals ejected in supernova explosions are mixed into the hot halo has. Chemical enrichment occurs when metal-rich gas from the halo cools and forms new disk stars. Note also that enrichment occurs more readily in lower mass galaxies, since the mass of gas that is returned to the intergalactic medium per solar mass of stars formed scales as $V_c^{-2}$ in equation 9, i.e gas can more easily escape the potential wells of less massive galaxies. As a result, disk galaxies in our model undergo rapid, early enrichment while they are still dwarf systems. One of the advantages of this chemical evolution scheme is that the disks then do not suffer from the classic "G-dwarf problem" that plagues simple closed-box models. This is illustrated in figure 3, where we plot the cumulative fraction of stars in the disk with metallicities smaller than a given value. As can be seen, only a few percent of the stars have metallicities less than 0.25 solar, in contrast to the closed box models where this fraction is almost always much higher.

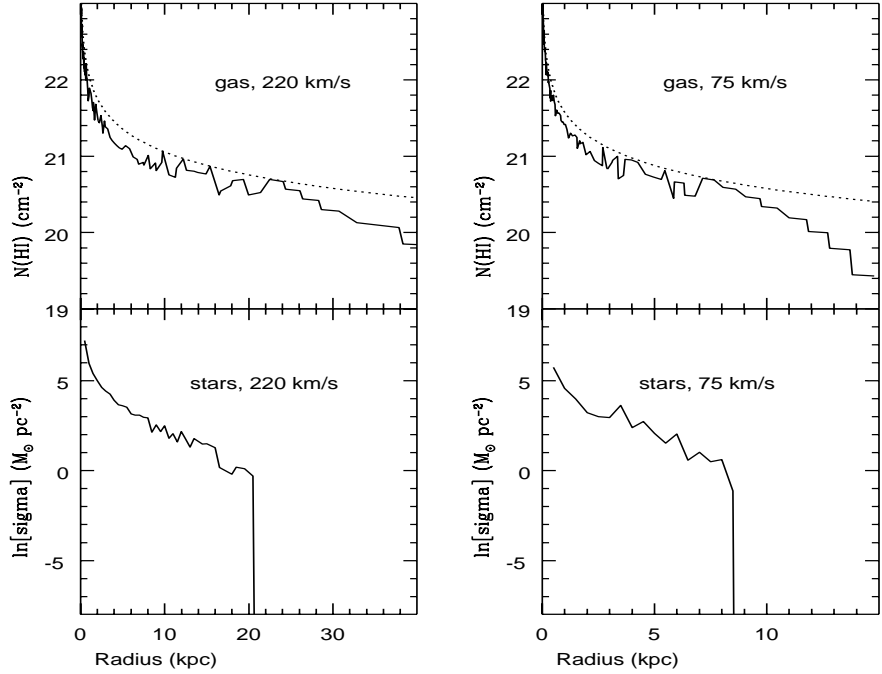

Figure 1: The gas and stellar surface density profiles of disk galaxies halos with circular velocity $V_c = 220$ km s$^{-1}$ and 75 km s$^{-1}$ at $z = 0$. The gas surface density is plotted in units of the HI column density seen through a disk with face-on orientation. The dotted line is the critical surface density. The stellar surface density is plotted in $M_\odot/pc^2$.

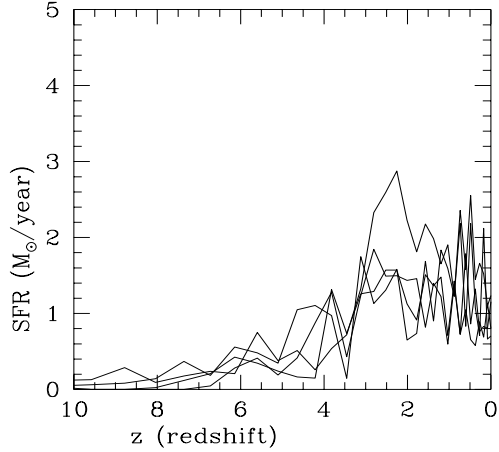

Figure 2: The star formation history of a disk galaxy that is in a halo with $V_c = 220$ km s$^{-1}$ at the present day. The different lines show various realizations of the formation history of the disk.

As noted by White & Frenk (1991), the chemical evolution scheme described above naturally results in a metallicity- luminosity relation in the sense that more luminous galaxies are more metal rich. There are two reasons for this effect. One is that large galaxies form stars for a longer period of time. The most important effect, however, is that a larger fraction of the gas reservoir in larger galaxies is turned into stars by the present day. The metallicity-luminosity relation that we obtain is shown in figure 4. It is in reasonable agreement with a fit to the observed relation from a compilation of data from different sources (Oey & Kennicutt (1993); Garnett & Shields (1987); Skillman et al (1989)) presented in a review article by Roberts & Haynes (1994).

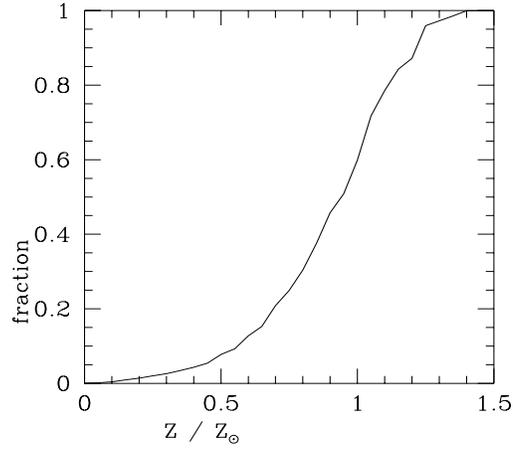

Figure 3: The percentage of stars in a Milky Way-type disk with metallicity less than $Z$

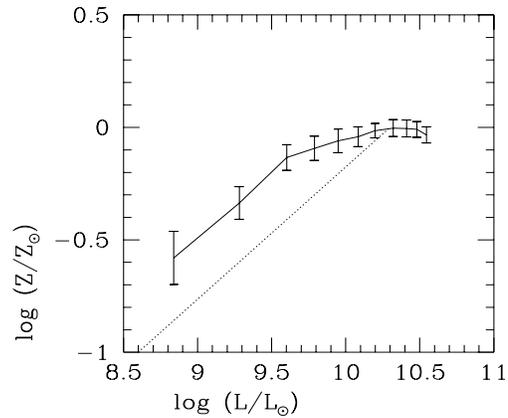

Figure 4: The mean metallicity of a galaxy (relative to solar) is plotted versus its B-band luminosity. The error bars indicate the scatter obtained for different disk formation histories. The dotted line is a fit to the observed relation from a compilation of data presented in a review article by Roberts & Haynes (1994)

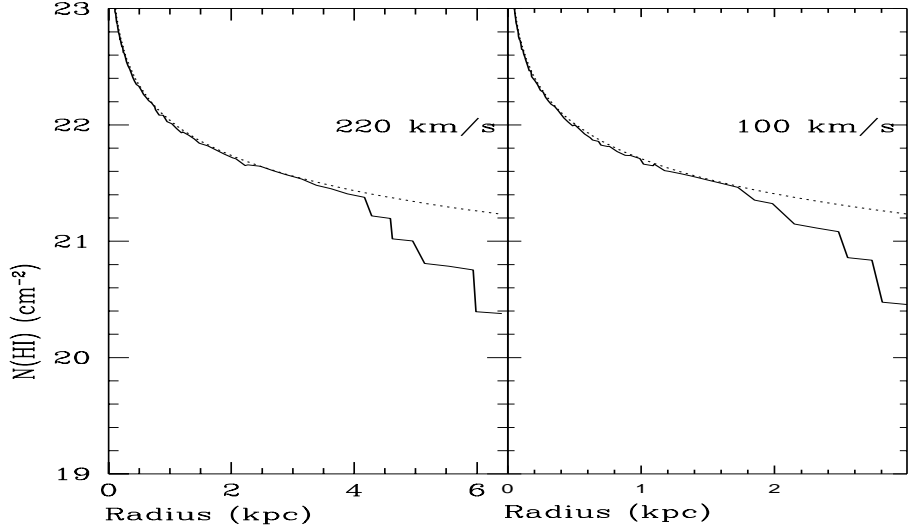

Figure 5: The gas surface density profiles of disk galaxies in halos of circular velocities 100 and 220 km s$^{-1}$ at a redshift $z = 2.5$. The dotted line shows the critical surface density.

## 3.2 Evolution of the Properties of Disk Galaxies at High Redshift

In the previous section, we demonstrated that our model is able to account successfully for many of the observed properties of disk galaxies at the present day. In this section, we extend our analysis to probe the properties of disk galaxies at high redshift.

The gas profiles of typical disk galaxies at z=2.5 are plotted in figure 5. The upper panel shows the profile of a disk with circular velocity 220 km s$^{-1}$ and the lower panel is for a disk with circular velocity 100 km s$^{-1}$. Since we assume that the disk formation process is the same at all redshifts, it is no surprise that the gas surface densities once again track the critical density over most of the disk. The main difference is that disks formed at higher redshift are smaller and more concentrated, since their virial radii scale as $(1+z)^{-3/2}$. The SFRs of high-redshift disks again range from a few tenths to a few solar masses per year. However, as will be seen later, at z=2.5 disks with circular velocities less than 100 km s$^{-1}$ dominate the total absorption crossection. These galaxies are inferred to have star formation rates of only a few tenths of a solar mass per year. It is therefore not a surprise that it has proved difficult to detect these objects in emission (see for example Hu et al 1993).

In order to calculate the *total* neutral hydrogen density $\Omega(HI)$ contributed by damped Ly$\alpha$ systems at a given redshift, one must know both the total amount of gas in each galaxy that contributes to the absorption, and the mass function of galaxies at that redshift. We will assume that the galaxies that contribute to damped Ly$\alpha$ absorption form at the centres of dark matter halos with circular velocities in the range 35 to 300 km s$^{-1}$. Figure 6 shows the contribution of galaxies of different circular velocities to the total crossectional area of gas at column densities greater than $2 \times 10^{20}$ cm$^{-2}$. Results have been plotted at four different redshifts. At z=0, galaxies with circular velocities greater than 100 km s$^{-1}$ make up about 50% of the total crossection; at z=2.5, this has decreased to 30% and by z=3.8, only 20% of the absorption area is produced by these more massive galaxies. In principle, this is something that can now be tested with kinematical data derived from high-resolution spectra (Wolfe et al 1994).

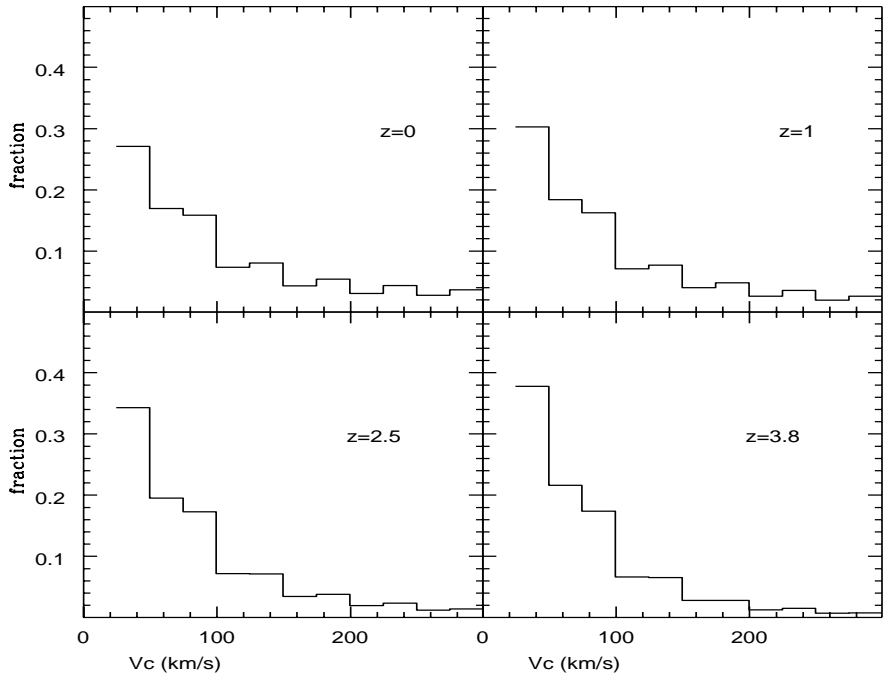

Figure 6: The contribution of galaxies of different circular velocities to the total crossectional area of HI gas in the universe with column densities greater than $2 \times 10^{20}$ cm$^{-2}$. Results are shown at four different redshifts.

In figure 7, we show the evolution of $\Omega$(HI) with redshift for a series of CDM models with different normalizations. The data points are from Storrie-Lombardi & MacMahon (1995) and are derived from a compilation of data from several different surveys. Their data includes 12 new high-redshift damped systems discovered in the APM QSO survey, together with all other existing lower redshift samples (Wolfe et al 1986; Lanzetta et al 1991; Lanzetta, Wolfe & Turnshek 1995). It should be noted that the new values of $\Omega$(HI) do not rise as steeply with increasing redshift as indicated by previous data. There appears to be evidence for a flattening at $z \sim 2$ and possibly even a turnover at $z \sim 3$. It is interesting that the same qualitative trends are apparent in the evolution of $\Omega$(HI) derived from the models. The most general conclusion is that the evolution predicted all the models is mild: $\Omega$(HI) increases by

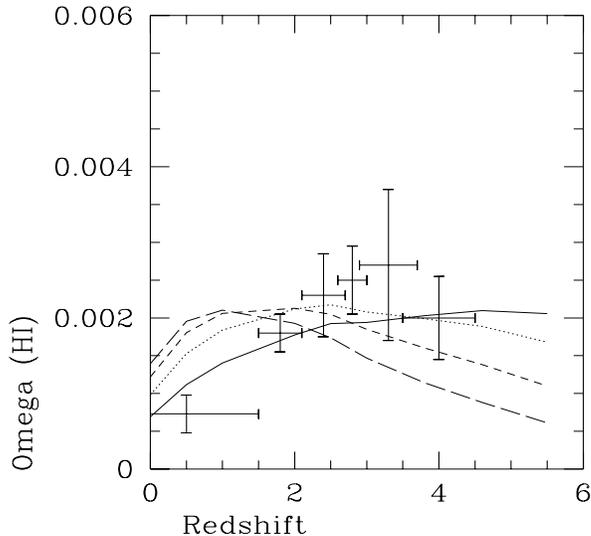

Figure 7: The evolution of $\Omega(HI)$ with redshift in CDM models with different normalization. The solid line is for $b = 1$, the dotted line $b = 1.5$, the short dashed line $b = 2$ and the long dashed line $b = 2.5$. The data points are taken from Storrie-Lombardi & MacMahon (1995). The value adopted for $H_0$ is 50 km s$^{-1}$ Mpc$^{-1}$.

*at most* a factor 3 from z=0 to z=3. This increase comes about because of the increase in the number of halos of galactic mass, and because of the shift in the distribution of galaxies to less luminous systems that are also more gas-rich. The redshift at which $\Omega$(HI) peaks depends on the normalization; models with high values of $b$ have late structure formation and $\Omega(HI)$ peaks at low redshift. It is clear, however, that much more data is needed before any constraints can be placed on cosmology.

Finally in figure 8, we show how the metallicity distribution of damped Ly$\alpha$ systems changes at high redshift. At z=0 the metallicity distribution is sharply peaked at values just under solar. By a redshift of 2.5. the metallicity distribution is much more evenly spread. The mean value is about 0.1 solar, but values as low as 0.01 solar and as high as 0.7 solar are expected. This accords rather well with the data of Pettini et al (1994), who find that the zinc abundances of damped systems at redshifts between 2 and 3 span a wide range. We also predict that at all redshifts, there should be a strong correlation between the metallicity of a damped systems and its circular velocity, with more metal-rich systems having higher rotation speeds. This is again something that can be checked using high resolution spectra.

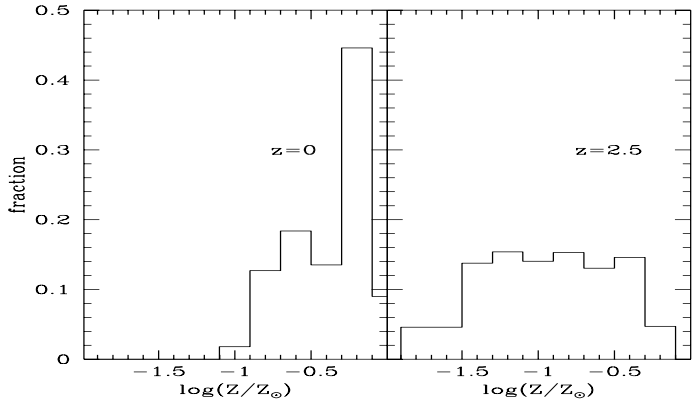

Figure 8: The metallicity distribution (relative to solar) of damped Ly$\alpha$ systems at $z = 0$ and at $z = 2.5$.

## 4  Discussion and Conclusions

The classical "closed box" approach to galaxy evolution is motivated by an Eggen, Lynden-Bell & Sandage-type picture of galaxy formation (1962), in which initially overdense regions in the early universe break away from the uniform Hubble expansion and then collapse monolithically into centrifugally supported gaseous disks. These disks then form stars over a Hubble time, becoming enriched in heavy elements in the process. The analysis in this paper is based on the White & Rees (1977) picture of galaxy formation, in which disk galaxies form by continuous cooling and accretion of gas within a merging hierarchy of dark matter halos. The Kennicutt law of star formation combined with this assumption of continuous infall results in gas in galactic disks being maintained at a critical threshold density $\Sigma_c$, where $\Sigma_c \propto V_c/R$. Chemical enrichment of the disks takes place when the surrounding hot halo gas is enriched in heavy elements ejected during supernova explosions. This gas then cools onto the disk producing a new generation of metal-rich stars.

We have shown that models of this type can reproduce many of the observed properties of present-day galactic disks like the Milky Way. These include gas and stellar surface density profiles, metallicities and the distribution of stars as a function of age and metal content. In particular, we find that the rapid early enrichment predicted by our model solves the classic G-dwarf problem.

We then use the models to make some inferences about the properties of disk galaxies at high redshift. Because the overall mass distribution in the universe shifts to smaller halos at higher redshifts, and these smaller halos contain less luminous, more gas-rich galaxies, we find that the total neutral hydrogen density $\Omega(HI)$ increases at higher z. The predicted increase, however, is rather mild, but is roughly consistent with the latest derivation of $\Omega(HI)$ as a function of z by Storrie-Lombardi & MacMahon (1995). Time will tell whether this model will still hold up when more data is accumulated. More extreme evolution would indicate the Kennicutt law does not hold for galaxies at high redshift and that some extra physical process must cause star formation to be less efficient at high z than at present. It is encouraging, however, that our models are also able to account for some of the other trends seen in the high-redshift data, including the increase in the number of high column-density systems at high redshifts, as well as the metallicity distribution of damped Ly$\alpha$ systems at $z \sim 2-3$.

Finally, one rather general prediction of all hierarchical models is that the galaxies that give rise to the damped Ly$\alpha$ absorption become progressively less luminous and more compact at higher redshift. This prediction will no doubt soon be tested by a new generation of telecopes and instruments capable of imaging galaxies as they were when the universe was young.